\documentstyle[11pt,baaa-eng,twoside,epsf,psfig]{article}
\begin{document}
\vskip 1.0cm
\markboth{T. E. Tecce, L. J. Pellizza \& A. E. Piatti}{Analysis of the chemical
evolution of the Galactic disk}
\pagestyle{myheadings}


\vspace*{0.5cm}
\parindent 0pt{ COMUNICACI\'ON DE TRABAJO -- CONTRIBUTED PAPER } 
\vskip 0.3cm
\title{Analysis of the chemical evolution of the Galactic disk via dynamical
simulations of the open cluster system} 

\author{Tom\'as E. Tecce}
\affil{Instituto de Astronom\'\i a y F\'\i sica del Espacio, CONICET / UBA, 
       Ciudad de Buenos Aires, Argentina, tomas@iafe.uba.ar}

\author{Leonardo J. Pellizza}
\affil{Service d'Astrophysique, CEA -- Saclay, Gif-sur-Yvette, Francia, 
leonardo.pellizza@cea.fr}

\author{Andr\'es E. Piatti}
\affil{Instituto de Astronom\'\i a y F\'\i sica del Espacio, CONICET / UBA, 
       Ciudad de Buenos Aires, Argentina, andres@iafe.uba.ar}

\begin{abstract} For several decades now, open clusters have been used to study
the structure and chemical evolution of the disk of our Galaxy. Due to the fact
that their ages and metallicities can be determined with relatively good 
precision, and since they can be observed even at great distances, they are 
excellent tracers of the variations in the abundance of heavy chemical elements
with age and position in the Galactic disk. In the present work we analyze the
star formation history and the chemical evolution of the disk of the Galaxy 
using numerical simulations of the dynamical evolution of the system of open 
clusters in the Milky Way. Starting from hypotheses on the history of cluster 
formation and the chemical enrichment of the disk, we model the present 
properties of the Galactic open cluster system. The comparison of these models 
with the observations allows us to examine the validity of the assumed 
hypotheses and to improve our knowledge about the initial conditions of the 
chemical evolution of the Galactic disk.
\end{abstract}

\begin{resumen} Desde hace ya varias d\'ecadas, los c\'umulos abiertos han sido
utilizados para estudiar la estructura y evoluci\'on qu\'\i mica del disco de 
la Galaxia. Debido a que sus edades y metalicidades pueden determinarse con 
precisi\'on relativamente buena, y a que pueden observarse incluso a grandes 
distancias, resultan excelentes trazadores de las variaciones de abundancia de
elementos pesados con la edad y la posici\'on en el disco gal\'actico. En el 
presente trabajo analizamos la historia de formaci\'on estelar y la evoluci\'on
qu\'\i mica del disco de la Galaxia utilizando simulaciones num\'ericas de la
evoluci\'on din\'amica del sistema de c\'umulos abiertos de la Galaxia. Para 
ello, a partir de hip\'otesis acerca de la historia de formaci\'on de c\'umulos
y del enriquecimiento qu\'\i mico del disco, mo\-de\-la\-mos las propiedades 
actuales del sistema de c\'umulos abiertos de la Galaxia. La comparaci\'on de 
dichos modelos con las observaciones nos permite examinar la validez de las 
hip\'otesis asumidas y mejorar nuestro conocimiento acerca de las condiciones 
iniciales de la evoluci\'on qu\'\i mica del disco gal\'actico. 
\end{resumen}

\section{Introduction}
The star formation history and chemical evolution of the galactic disk are
still a subject of debate. The evolution of the star formation rate (SFR) of 
the disk from its formation to the present is not well known. Some authors 
suggest that a constant SFR could explain the observations (e.g., Twarog 1980),
while others claim that it consisted of a series of isolated bursts (e.g., 
Rocha-Pinto et~al. 2000; Lamers et~al. 2005). The chemical homogeneity of the 
galactic disk and the possible existence of an age-metallicity relationship 
(AMR) are also being discussed. Previous works have measured a radial 
metallicity gradient in the disk (e.g. Piatti, Clari\'a \& Abadi 1995; Chen, 
Hou \& Wang 2003, hereafter CHW), but failed to establish beyond any doubt the 
existence of a vertical gradient. The possible time variation of such gradients
is also unknown. Finally, the existence of an AMR for disk stars is claimed by 
some authors (e.g., Rocha-Pinto et~al. 2000), but denied by others (e.g., 
Feltzing et~al. 2001).

One approach to these problems that has proven to be fruitful is the use of
open clusters (OCs) to trace the SFR and chemical abundances in the galactic
disk. Some of the quoted results have been obtained using this approach, either
by directly determining the properties of the disk (gradients, AMR) from the
observed sample of OCs (e.g. CHW), or by tracing the motion of a few clusters 
back in time to study the evolution of these properties (Carraro \& Chiosi 
1994; Piatti, Clari\'a \& Abadi 1995). Although this approach takes advantage 
of the precision with which OC positions, ages and metallicities can be 
measured, its results are affected by the incompleteness and inhomogeneity of 
the available samples of OCs.

In this paper we present an alternative approach to investigate the problem of 
the SFR and chemical evolution of the galactic disk. We simulate numerically 
the origin and evolution of the OC system from the formation of the disk up to 
the present time. We assume particular models for the star formation history of
the disk and the relationship between star formation and cluster formation to 
create a set of simulated OCs. Metallicities of these OCs are assigned 
according to models for the chemical evolution of the disk. Using a standard 
galactic model and a cluster destruction model, we compute the evolution of 
this system to simulate the properties of the {\em present} OC system. Finally,
simulating realistic observational selection effects we construct samples of 
clusters similar to those observed, to which they are compared. Our approach 
shares the advantages of previous ones, the use of OCs gives us a sample of 
objects with precise positions, ages and metallicities, and the inclusion of 
dynamical evolution allows us to consider the evolution of disk properties. On 
the other hand, the simulation of the whole OC system and of the observational 
selection effects gives this approach the additional advantage of improving the
comparison of the models with observations.

\section{Simulations}
Our simulations are based on the standard galactic model of Dehnen \& Binney
(1998), which describes the mass density $\rho_m$ and gravitational
potential of the various subsystems of the Milky Way. We use 10~Gyr ago as
the initial time, assuming that the disk was already formed at that time and
that the Galaxy did not undergo considerable changes since then, so that the
model remains a reasonable approximation to its structure throughout the whole
simulated time interval. The star formation history of the galactic disk is
described by its SFR per unit volume $\rho_{\mathrm{SFR}}$, whose spatial
dependence is assumed to be given by the Schmidt law (e.g., Kennicutt 1998),
\begin{equation}
\rho_{\mathrm{SFR}}(\vec{x},t) = \rho_{\mathrm{gas}}^{1.5}(\vec{x}) f_1(t),
\end{equation}
where $\vec{x}$ is the position in the galaxy, $t$ the look-back time and
$f_1(t)$ is proportional to the SFR, and is left free to explore different
models of the latter. The chemical evolution of the disk is modeled by a
function $\mu(\vec{x},t)$ which gives the age-metallicity-position
relationship, and is also left free to explore different possibilities.

The dimension of a typical OC ($\sim$10~pc) is much smaller that the typical
scale length in which the galactic potential changes, hence OCs can be
described as point objects in our simulations. The OC system is then
represented by a set of $N$ particles, each of them described by a birth time,
position, velocity, mass and metallicity [Fe/H]. To generate the $N$ particles,
we used a Monte Carlo method with probability distributions given by the model
described above. We assumed that a constant fraction of the stars are formed
in clusters, and also a constant cluster initial mass function (CIMF). In this
case, the cluster formation rate per unit volume is proportional to the SFR per
unit volume, and the function $\rho_{\mathrm{SFR}}(\vec{x},t)$ gives the 
probability distribution for birth times and positions. The cluster velocities 
were assumed to have a Gaussian distribution, with a mean given by the local 
circular velocity at the place of birth, and a dispersion of 10~km~s$^{-1}$, 
which corresponds to the velocity dispersion of the giant molecular clouds 
that give birth to the OCs (Dickey \& Lockman 1990). For the masses $m$ of the
OCs we used as a probability distribution the CIMF, $\psi(m) \propto m^{-2}$ 
(Lamers et~al. 2005), while $\mu(\vec{x},t)$ gave us the [Fe/H] metallicity 
probability distribution at position $\vec{x}$ and look-back time $t$.

Each cluster was given a lifetime according to the OC destruction model of
Lamers et al. (2005), and discarded from the sample if its destruction
occurred before the present time. Dynamical evolution for each surviving 
cluster was computed from its time of birth to the present time by integrating 
its equations of motion in the galactic potential given by the model of Dehnen 
\& Binney (1998). In this way we obtained a set of surviving clusters which
represents the {\em present} OC system of the Milky Way.

\section{Results}
In the first simulation we assumed $f_1(t) = K$, with $K$ a constant; in this 
simulation we disregarded the metallicities. We selected from the resulting OC 
system the sample of all clusters within 600~pc from the Sun, to compare its 
age distribution with that of the OC catalogue from Kharchenko et~al. (2005), 
which is complete up to this distance. A total of $N = 7.5 \times 10^6$ 
clusters were simulated, which results in an average of 100 clusters within 
600~pc from the Sun, a number comparable to that in the Kharchenko et al. 
(2005) catalogue. The comparison is shown in Figure 1, and it suggests that the
SFR was indeed almost constant during the whole evolution of the disk, except 
for some brief events. A high SFR period between 0.60 and 0.25~Gyr ago followed
by a low SFR period between 0.16 and 0.06~Gyr ago are the most prominent 
events. A constant SFR does not fit the data in these age intervals, even 
taking into account observational errors. These results agree with those of 
Lamers et~al. (2005) and de La Fuente Marcos \& de La Fuente Marcos (2004). On 
the other hand, the comparison of the total number of observed and simulated 
clusters implies a total present number of clusters in the Galaxy $N_0 = (2.2 
\pm 0.3) \times 10^5$, of the same order of magnitude of that obtained by 
Piskunov et~al. (2005), and a total SFR in the disk of 
$(0.9 \pm 0.1)$~$M_\odot$~yr$^{-1}$, consistent with the values for normal 
spiral galaxies (Kennicutt 1998).

\begin{figure}  
{\hspace*{2.5cm}   \psfig{figure=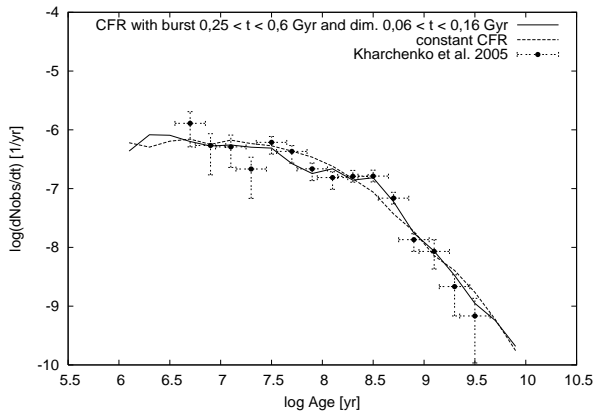,width=8.cm}}
\caption{Age histogram in units of number of clusters per time interval, in 
logarithmic age bins of 0.2. Points with error bars are data for 114 OCs with
$d < 600$~pc from the catalogue of Kharchenko et al. (2005). The lines 
correspond to the results of our simulations with a constant CFR (dotted line)
and a CFR with a burst between 0.25 and 0.60~Gyr ago, and a low SFR period 
between 0.06 and 0.16~Gyr ago.}
\end{figure}

In our second simulation we used the function $f_1(t)$ obtained from the
comparison made in the first one and assumed no AMR, that is, a random
metallicity distribution in the interval $[-0.7, 0.3]$. From the resulting 
system we selected all the OCs with 5~kpc~$< R <$~14~kpc, a sample comparable 
to the catalogue of CHW. We observe in Figure 2(a) that both samples clearly 
disagree. Although the sample of CHW is not complete, any selection effects 
invoked to restore the agreement between simulations and observations would 
link metallicities either to positions or ages, contradicting the hypothesis of
random metallicities. Hence we conclude that this hypothesis does not produce a
good model of the chemical evolution of the galactic disk.

Our third simulation changed the metallicity model to an homogeneous disk with 
a simple AMR with constant slope, particularly that proposed by Rocha-Pinto et~al. (2000),
\begin{equation}
\mu(\vec{x},t) = 0.2 - 0.09 \,\mathrm{Gyr}^{-1} t,
\end{equation}
plus a random component of zero mean and dispersion equal to $0.15$~dex. Using
the same selection process as in the second simulation, we obtained a sample
of OCs to compare with that of CHW. Figure 2(b) shows a good agreement between 
both samples. Nevertheless, the simulated sample shows no radial metallicity 
gradient, while the existence of such a gradient is well established. 
Furthermore, any selection effects invoked to restore agreement would imply a
relation between metallicity and position in the disk, thus contradicting the
initial assumption of chemical homogeneity. Thus we also discarded this 
chemical evolution model.

\begin{figure}
\hbox{ 
   \psfig{figure=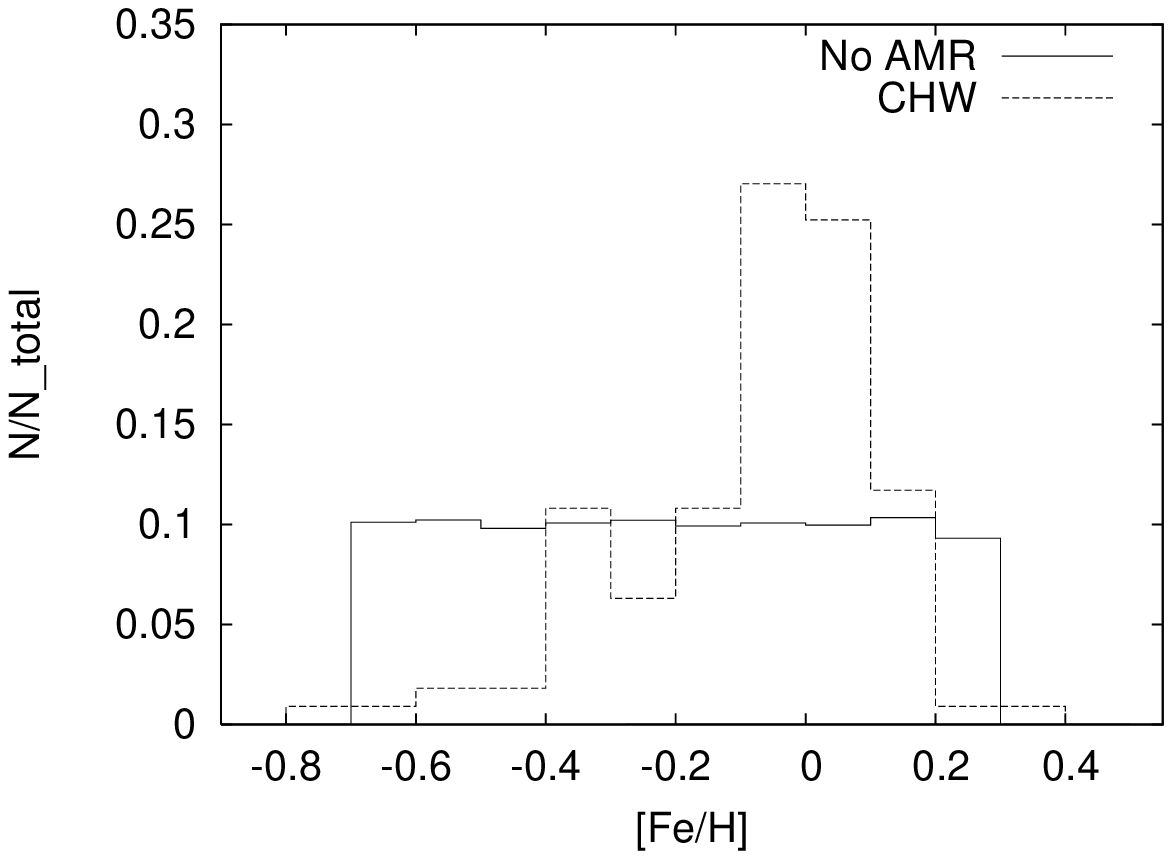,width=6.cm}
          \hspace*{0.2cm}
   \psfig{figure=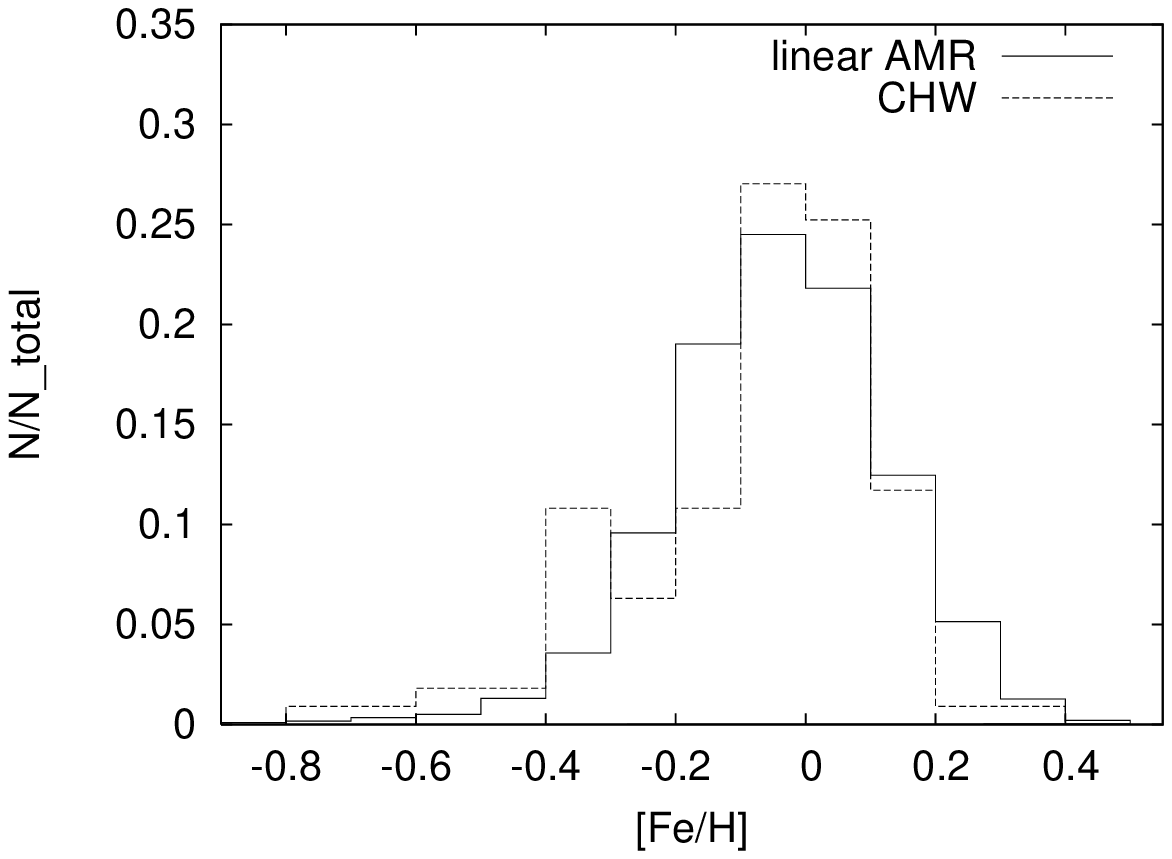,width=6.cm}
        }
\caption{{\it (a)} Metallicity distribution obtained when no AMR is assumed, 
compared with the distribution obtained using the clusters of the CHW 
catalogue. {\it (b):} Metallicity distribution obtained using the AMR proposed 
by Rocha-Pinto et al. (2000), compared with the distribution obtained using the
CHW catalogue.}
\end{figure}

The third simulation has shown that the radial metallicity gradient cannot be
created by the dynamical evolution, hence it must be present in the gas from
which OCs are created. We performed a fourth simulation, with a different
chemical model in which we included a radial gradient but no AMR. The
metallicity in this model is given by
\begin{equation}\label{grad}
\mu(\vec{x},t) = 0.75 - 0.09 \,\mathrm{kpc}^{-1} R.
\end{equation}
The slope of the gradient was selected to be the same gradient observed in the
OC system at the present (Parisi et al. 2005). A change in the slope does not
depend on its particular value, and if there is no change this choice would 
result in a self-consistent model. In this case, the OC sample was selected in 
the same way as before, but a selection effect was simulated assuming that the 
sample of CHW is limited in magnitude and computing the apparent magnitude of 
each cluster from its mass (assuming as a representative value a mass-to-light 
ratio of 1~$M_\odot / L_\odot$) and the extinction derived from the hydrogen 
column density, calculated using the ISM density given by the model of Dehnen 
\& Binney (1998). Because of this effect, highly absorbed OCs are undetectable.
These clusters are mainly located towards the Galactic center; this, combined 
with the radial gradient, cuts off the high metallicity part of the histogram. 
The results are shown in Figure 3, where the results of the simulation are 
fitted to equation \ref{grad} via linear least squares, showing a very good 
agreement between both samples. This simulation also shows that the radial 
metallicity gradient is preserved during the whole dynamical evolution of the 
disk. Hence, the results support the model proposed in this simulation for the 
chemical evolution of the Galactic disk.

\begin{figure}  
{\hspace*{2.5cm}   \psfig{figure=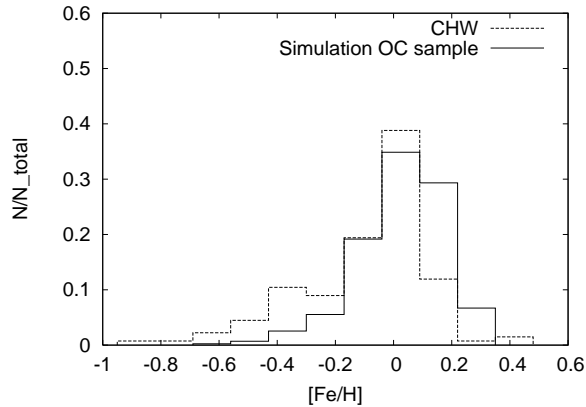,width=8.cm}}
\caption{Histogram of the metallicity distribution obtained simulating a 
magnitude limit selection effect, compared with the distribution obtained from
the clusters in the CHW catalogue.}
\end{figure}

\section{Conclusions}
We performed numerical simulations of the OC system of the Milky Way to test
different SF history and chemical enrichment models of the disk. The results
of our simulations suggest that the SFR of the disk has been practically
constant in the last 10~Gyr, with only a few time intervals of enhanced or
lowered SFR. Our results show also that the simplest way to explain both the
observed metallicity distribution and radial metallicity gradient of the
observed clusters is to assume that the latter is primordial and did not
change during the evolution of the disk, and that there is no age-metallicity
relationship in the disk. Nevertheless, the existence of both an AMR and a 
radial metallicity gradient is not ruled out by our models.

\end{document}